\documentclass{article}

\usepackage{arxiv}

\usepackage[utf8]{inputenc} % allow utf-8 input
\usepackage[T1]{fontenc}    % use 8-bit T1 fonts
\usepackage{hyperref}       % hyperlinks
\usepackage{booktabs}       % professional-quality tables
\usepackage{nicefrac}       % compact symbols for 1/2, etc.
\usepackage{microtype}      % microtypography
\usepackage{doi}

\usepackage{amsmath,amssymb,amsfonts,amsthm, mathtools}
\usepackage{bm}
\usepackage{graphicx}
\usepackage{subfig}
\usepackage{url}
\usepackage{hyperref}
\usepackage{algorithm}
\usepackage{algorithmic}
\usepackage{soul, xcolor}
\DeclarePairedDelimiter{\nint}\lfloor\rceil

\title{Modal Estimation on a Warped Frequency Axis for Linear System Modeling}

\author{\href{https://orcid.org/0000-0002-3890-6604}{\includegraphics[scale=0.06]{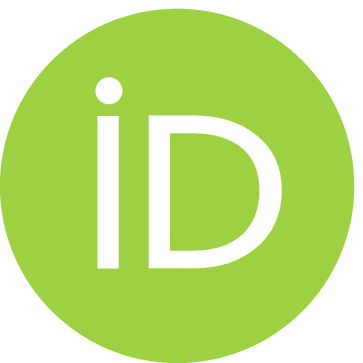}\hspace{1mm}Orchisama Das}\thanks{Use footnote for providing further
		information about author (webpage, alternative
		address)---\emph{not} for acknowledging funding agencies.} \\
	Institute of Sound Recording,\\
	University of Surrey\\
	Guildford, GU2 7XH, UK \\
	\texttt{o.das@surrey.ac.uk} \\
	\And
    {Jonathan S. Abel} \\
	Center for Computer Research in Music and Acoustics\\
	Stanford University\\
	Stanford, CA 94305, USA \\
	\texttt{abel@ccrma.stanford.edu} \\
}

\date{}

\renewcommand{\shorttitle}{\textit{arXiv} Template}

\hypersetup{
pdftitle={Modal Estimation on a Warped Frequency Axis for Linear System Modeling},
pdfsubject={q-bio.NC, q-bio.QM},
pdfauthor={Orchisama Das, Jonathan S. Abel},
pdfkeywords={System Identification, Frequency Warping, ESPRIT},
}

\begin{document}
\maketitle

\begin{abstract}
Linear systems such as room acoustics and string oscillations may be modeled as the sum of mode responses, each characterized by a frequency, damping and amplitude. Here, we consider finding the mode parameters from impulse response measurements, and estimate the mode frequencies and decay rates as the generalized eigenvalues of Hankel matrices of system response samples, similar to ESPRIT. For greater resolution at low frequencies, such as desired in room acoustics and musical instrument modeling, the estimation is done on a warped frequency axis. The approach has the benefit of selecting the number of modes to achieve a desired fidelity to the measured impulse response.  An optimization to further refine the frequency and damping parameters is presented.  The method is used to model coupled piano strings and room impulse responses, with its performance comparing favorably to FZ-ARMA.
\end{abstract}

% keywords can be removed

%%%%%%%%%%%%%%%%%%%%%%%%%%%%%%%%%%%%%%%%%%%%%%%%%%%%%%%%%%%%%%%%%
\section{Introduction}
Modal structures are an efficient way to synthesize acoustic spaces and vibrating systems. Modes originate from standing waves --- they are exponentially damped sinusoids that are characterized by their frequencies, amplitudes and decay rates. These damped sinusoids are eigenfunctions of the acoustic transfer function between the input and output pressure waves. Modal analysis aims to estimate these parameters, and modal synthesis resynthesizes the analyzed sound using a bank of parallel biquad filters, each implementing one mode \cite{adrien1991missing}.

In \cite{abel2014modal}, mode frequencies of a room are estimated by finding peaks in the power spectrum of the impulse response, and decay rates are approximated using reverberation times in bands about the estimated mode frequencies. Since rooms have thousands of modes, it is common-practice to estimate them on a band-by-band basis \cite{abel2014modal,maestre2017constrained}. Mode frequencies of carillon bells are estimated similarly in \cite{rau2019improved}, and the decay rates are found with non-linear optimization \cite{antsalo2001estimation}. Frequency-zoomed ARMA modeling on filtered groups of resonant frequencies has been used to model noisy string instruments and room responses in \cite{karjalainen2002frequency}. In \cite{abel2020direct,kereliuk2018Modal}, modal parameters are estimated on a subband basis from the generalized eigenvalues of shifted Hankel matrices formed with impulse response samples. The method in \cite{kereliuk2018Modal} is computationally identical to ESPRIT \cite{roy1989esprit}, and was first formalized by Hua and Sarkar in \cite{hua1990matrix}. In \cite{kereliuk2018Modal}, the number of modes in each band is determined by k-means clustering, whereas in \cite{abel2020direct}, they are determined by the relative magnitude of the singular values of the Hankel matrix. Modal parameter estimation with ESPRIT for modeling impulse responses has also recently been explored in \cite{wells2021modal}.

Frequency warping \cite{harma2000frequency} replaces the unit digital delay operator with a first order allpass filter with a non-uniform group delay. A cascade of allpass sections introduces non-uniform group delay, where the low-frequency components are delayed in time and the high frequency components get advanced. A warped frequency axis emphasizes psychoacoustic perception with more resolution in low frequencies. Warped digital filters have applications in loudspeaker equalization, linear predictive coding and physical modeling. Warping an impulse response has the effect of spreading out low-frequency modes around the unit circle, so that beating frequencies can be separated and resolved. Resolution of closely spaced sinusoids is important in modeling coupled vibrations. Some methods to address this problem include subspace methods such as \cite{das2018fast}, or spectral windowing methods such as the one used in \cite{rau2019improved}. However these methods require parameter tuning or are computationally expensive. In this paper, we perform modal estimation on a warped impulse response to resolve low-frequency beating modes.

Once the mode frequencies and dampings are estimated from the warped signal, we unwarp and optimize them. A sequential time-domain optimization scheme inspired by \cite{ainsleigh1992modeling} is used. The mode amplitudes are re-estimated using least squares in each iteration. This kind of sequential optimization with consecutive update of the linear and non-linear parameters has also been used in \cite{das2021room}. This is different from  \cite{maestre2017constrained,maestre2016design} where Maestre et al. fine-tune initial estimates of mode parameters with frequency-domain pole optimization.

This paper is organized as follows. We first discuss modal parameter estimation with ESPRIT in Section~\ref{sec:modes} and frequency-zoomed modal estimation with subband ESPRIT in Section~\ref{sec:freq_zoom}. We introduce the frequency warped modal estimator in Section~\ref{sec:warp}, and propose a new time-domain mode optimization method in Section~\ref{sec:opt}. In Section~\ref{sec:appli}, we discuss two applications of the proposed method --- modal estimation of coupled piano strings (Section ~\ref{ssec:coupled_piano}) and room impulse responses (Section~\ref{ssec:rir}). Each piano key (except the lower keys) is associated with sets of two or three strings coupled at the bridge. These strings are tuned with a small frequency deviation. String coupling leads to two-stage decay and beating \cite{weinreich1977coupled,weinreich1979coupled}. Room impulse responses also have a dense distribution of closely-spaced modes, some of which decay rapidly. Finally, we discuss the advantages of our method in Section~\ref{sec:disc} and conclude the paper in Section~\ref{sec:conclusion}.

%%%%%%%%%%%%%%%%%%%%%%%%%%%%%%%%%%%%%%%%%%%%%%%%%%%%%%%%%%%%%%%%

\section{Modal Estimation}
\label{sec:modes}

When a string is struck or plucked or a room is excited, traveling waves move in opposite directions, get reflected at the bridge or walls, and keep traveling back and forth. The resultant motion creates standing waves. The standing waves dissipate energy with time because of scattering and absorption. These standing waves, or modes of vibration, are damped sinusoids vibrating at natural frequencies of the system. Diagonalizing the second order partial differential equation of the one-dimensional traveling wave on a string, or three-dimensional traveling wave in a room yields damped sinusoids as the eigenfunctions of the system. The resulting vibration is due to the combination of several modes. 
% Two strings that are coupled at a bridge have two normal modes of vibration, the \textit{symmetric} mode causing ``in-phase'' vibrations and the \textit{anti-symmetric} mode causing ``anti-phase'' vibrations.  %to-do add paragraph about damped sinusoids being the eigenmodes of the 1D wave PDE.

A discrete-time modeled signal $\hat{h}(t), t = 0, 1, \ldots, T$, represented by a rational system without repeated poles, can be written as a sum of $M$ modes,
\begin{equation}
    \begin{split}
    \label{eq:modes}
    \hat{h}(t) = \sum_{m=1}^M  \gamma_m e^{( j\omega_m - \alpha_m) t } 
    % h_m(t) &= \gamma_m e^{( j\omega_m - \alpha_m) t }
    \end{split}
\end{equation}
where $\omega_m$ is the angular frequency, $\alpha_m$ is the decay rate and $\gamma_m$ is the complex amplitude of the $m$th mode. The goal is to estimate the mode parameters from a noisy measurement of the signal, $h(t)$. The modal reconstruction filter is formed with a parallel bank of $M$ second order filters, each synthesizing one mode. The Z-domain representation of the modeled signal $\hat{H}(z)$ is a sum of biquad filter transfer functions, i.e.,
\begin{equation}
\label{eq:resonant_biquad}
    \begin{split}
        \hat{H}(z) &= \mathcal{Z}\left(\sum_{m=1}^M \Re(h_m(t))\right) \\
        &= \sum_{m=1}^M \frac{\Re(\gamma_m) - e^{-\alpha_m}\Re(\gamma_m e^{-j\omega_m})z^{-1}}{1 - 2e^{-\alpha_m}\cos{\omega_m}z^{-1} + e^{-2\alpha_m}z^{-2}}
    \end{split}
\end{equation}
\noindent where $\Re$ denotes the real part of a complex number.

Consider a Vandermonde matrix, $\bm{V}$, with the  $m$th column, $\bm{v}_m$, representing the time series of the $m$th mode, $h_m(t)$

\begin{equation}
    \begin{split}
        \bm{V} &=\begin{bmatrix} \bm{v}_1 & \bm{v}_2 & \ldots & \bm{v}_M \end{bmatrix} \\
        \bm{v}_m &= \begin{bmatrix} e^{-\left(j\omega_m - \alpha_m\right)0} & \cdots 
        & e^{-\left(j\omega_m - \alpha_m\right)\frac{T}{2}} \end{bmatrix}^\mathsf{H}
    \end{split}
\end{equation}
where $\mathsf{H}$ denotes Hermitian transpose.
The Hankel matrix formed by the signal samples, $\bm{H}$, can be written as an outer product of the Vandermonde matrix $\bm{V}$ with a diagonal matrix of mode amplitudes, $\bm{\Gamma} = \text{diag}\left[\gamma_1, \ldots, \gamma_M \right]$.

\begin{gather}
    \bm{H} = \begin{bmatrix}
    h(0) & h(1)  & \ldots & h(\frac{T}{2}) \\
    h(1) & h(2) &  \ldots & h(\frac{T}{2}+1) \\
    \vdots &  \vdots & \ddots & \vdots \\
    h(\frac{T}{2}) & h(\frac{T}{2}+1) & \ldots & h(T) 
    \end{bmatrix} \\
    \bm{H} = \bm{V \Gamma V}^\mathsf{H}
\end{gather}

\noindent Similar to ESPRIT \cite{roy1989esprit}, the Hankel matrix offset by $1$ sample, $\bm{K}$, can be written as

\begin{gather}
% \begin{split}
    \bm{K} = \begin{bmatrix}
    h(1) & h(2)  & \ldots & h(\frac{T}{2}+1) \\
    h(2) & h(3) &  \ldots & h(\frac{T}{2}+2) \\
    \vdots &  \vdots & \ddots & \vdots \\
    h(\frac{T}{2}+1) & h(\frac{T}{2}+2) & \ldots & h(T+1) 
    \end{bmatrix} \\
    \bm{K} = \bm{V \Psi \Gamma V}^\mathsf{H}
    % \end{split}
\end{gather}
where $\bm{\Psi} = \text{diag}\left[e^{\left(j\omega_1 - \alpha_1 \right)} \ \ldots \ e^{\left(j\omega_M - \alpha_M \right)}\right]$. Post multiplying $\bm{K}$ with the pseudoinverse of $\bm{H}$,  we get
\begin{equation}
    \bm{K H}^\dagger = \bm{V \Psi V}^{-1}
\end{equation}
Thus, the diagonal elements of $\bm{\Psi}$ are the generalized eigenvalues of the matrix pencil $\left(\bm{K},\bm{H}\right)$. The mode frequency and damping estimates are the imaginary and real parts of the logarithm of the eigenvalues, $\psi_m$,
\begin{equation}
\label{eq:log_eig}
    \ln \psi_m = j\omega_m - \alpha_m 
\end{equation}
The number of modes, $M$, is estimated by the rank of the matrix $\bm{H}$ (by its largest singular values). The mode amplitudes, $\bm{\gamma}$, are found by least squares fit to the measured signal.
\begin{equation}
    \bm{\gamma} = \bm{V}^\dagger\bm{h}
\end{equation}
where $\bm{V}^\dagger$ is the pseudoinverse of the Vandermonde matrix formed from the mode frequencies and dampings.

%%%%%%%%%%%%%%%%%%%%%%%%%%%%%%%%%%%%%%%%%%%%%%%%%%%%%%%%%%%%%%%%%%

%%%%%%%%%%%%%%%%%%%%%%%%%%%%%%%%%%%%%%%%%%%%%%%%%
\section{Frequency-zoomed Modeling}
\label{sec:freq_zoom}

The modal estimation method described in Section~{\ref{sec:modes}} can be done on a frequency band-by-band basis. The signal, $h(t)$, is filtered into overlapping frequency bands, and downsampled. This process ensures that each band has relatively few modes to be estimated that are well separated in frequency. In other words, subband processing puts a magnifying glass on the frequency spectrum, and helps resolve close frequency beating modes. Furthermore, downsampling effectively increases the pole damping.

Let us consider $N_b$ bands at a sampling rate of $f_s$~Hz, each with band center at $f_n$~Hz, $f_n = (2n-1)f_s/4N_b$, $n = 1,2, \cdots, N_b$. The impulse response is heterodyned by multiplying with a complex exponential tuned to the band centers,
\begin{equation}
    h_n(t) = h(t)  \exp{(-2\pi j f_n t)}.
\end{equation}
The heterodyned signal, $h_n(t)$, which is now centered at DC, is then filtered with a high-order LPF with a narrow passband. The filtered signal is decimated by a factor of $r$. The mode frequencies and dampings are estimated for each band as $\hat{f}_{m,n}$ and $\hat{\alpha}_{m,n}$. Mode frequencies are moved up the spectrum to $\hat{f}_{m,n} + f_n$ to undo the effect of heterodyne,  and mode amplitudes are adjusted as $\sqrt[r]{\hat{\alpha}_{m,n}}$ to undo the effect of downsampling. Modes from subsequent bands are concatenated after discarding modes in overlapping passbands. 

This process is comparable to FZ-ARMA \cite{karjalainen2002frequency} where the signal is also processed in frequency bands, heterodyned and downsampled before detecting ARMA model coefficients with Steiglitz-McBride iteration. Therefore, we will refer to this method as FZ-ESPRIT (Frequency-Zoomed ESPRIT) henceforth in this paper. One advantage of modeling the system as second order filters in parallel versus a large order polynomial, as is the case in ARMA models, is that numerical methods such as Steiglitz McBridge can be unstable with high order models, whereas ESPRIT is numerically robust and gives stable solutions. Moreover, model order selection in FZ-ARMA is not straightforward, whereas in FZ-ESPRIT, the number of modes in any frequency band are picked based on the magnitude of the singular values of the Hankel matrix. 

%%%%%%%%%%%%%%%%%%%%%%%%%%%%%%%%%%%%%%%%%%
\section{Proposed Method}

\subsection{Frequency warped modal estimation}
\label{sec:warp}

\begin{figure}[t]
    \centering
     \subfloat[][Warping an impulse response]{
    \includegraphics[trim={0 7.8in 0 0},clip]{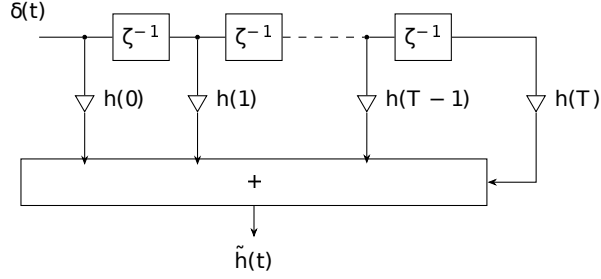} 
    \label{fig:tikz_warp}
    }\\
    \subfloat[][Frequency warping as conformal mapping]{
    \includegraphics[height=0.2\textheight]{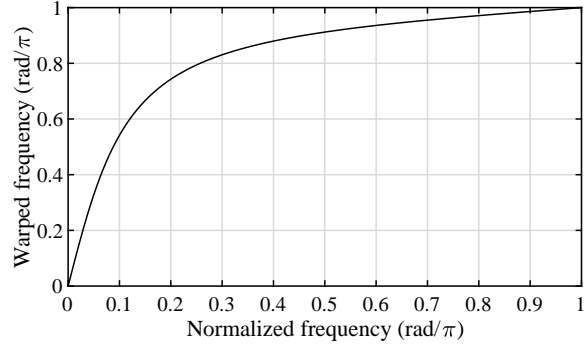} 
    \label{fig:map}
    }
    \caption{Frequency warping with allpass sections.}
\end{figure}

\begin{figure*}
    \centering
    \subfloat[][Original signal]{
    \includegraphics[height=0.25\textheight]{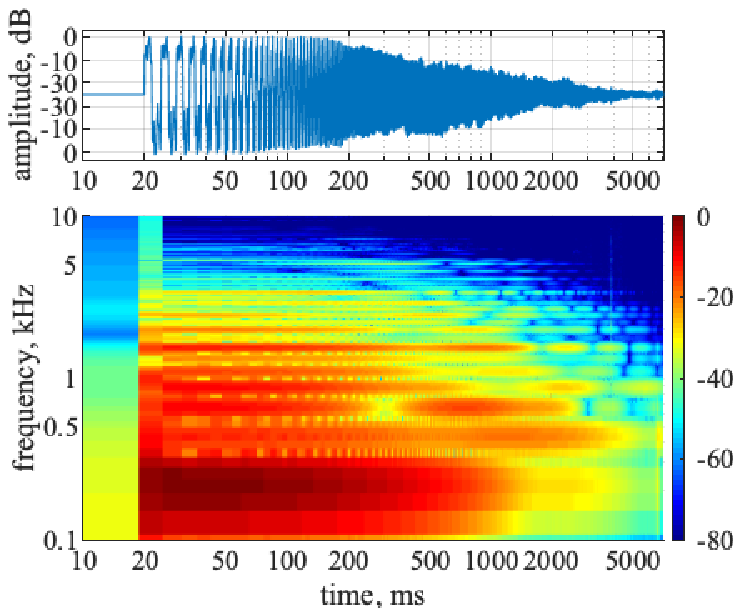}
    \label{subfig:ir}
    } \qquad
    \subfloat[][Warped Signal]{
    \includegraphics[height=0.25\textheight]{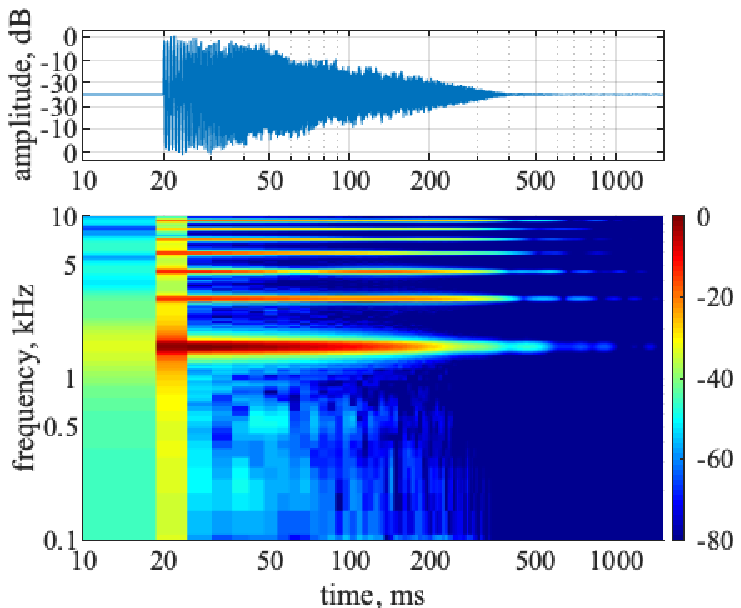}
    \label{subfig:irwarp}
    } \\
    \vspace{0.2in}
    \subfloat[Modes on unit circle without warping]{
    \includegraphics[height=0.4\textheight]{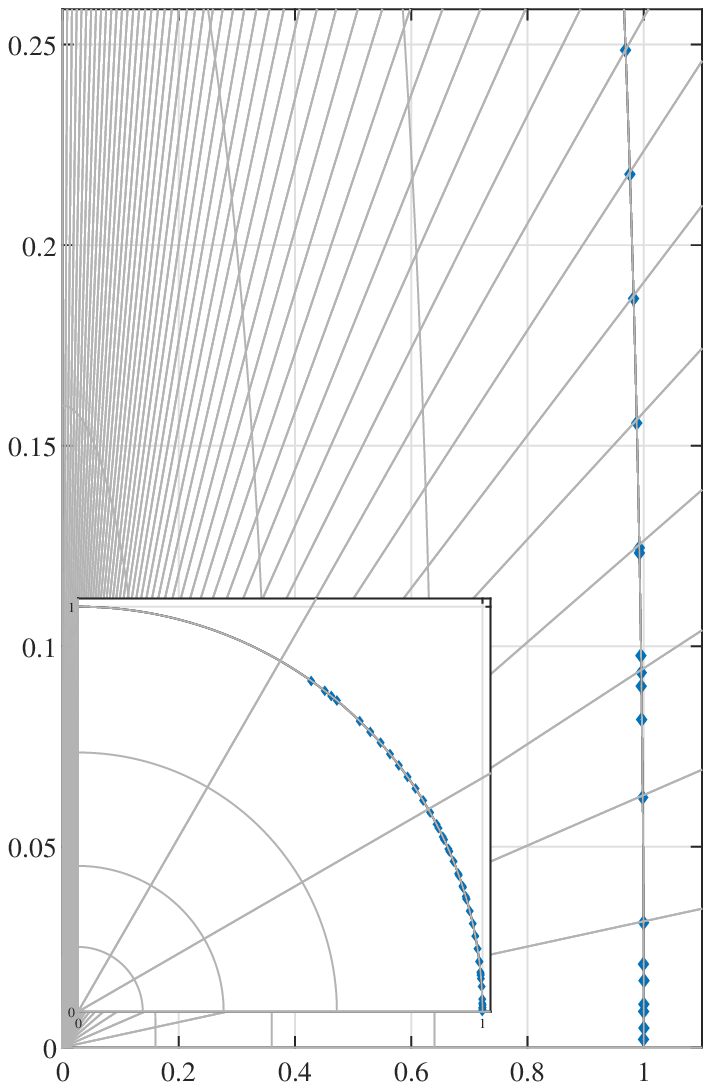}
    \label{subfig:no_warp_poles}
    }\qquad
    \subfloat[Modes on unit circle with warping]{
    \includegraphics[height=0.4\textheight]{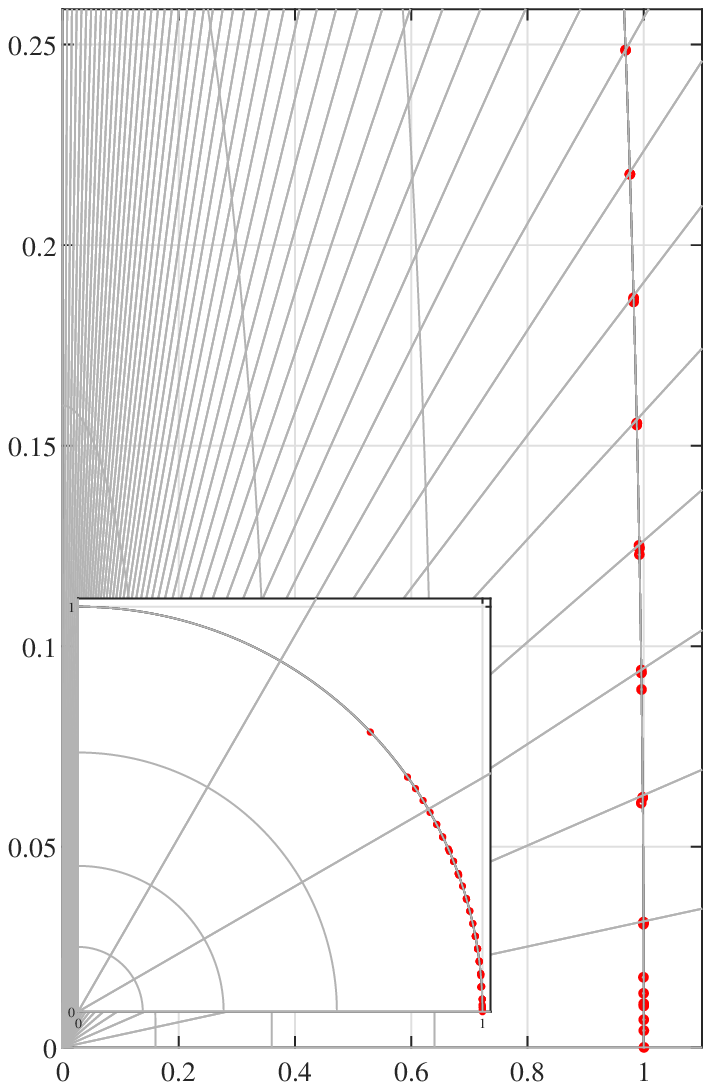}
    \label{subfig:warp_poles}
    }
    \vspace{0.1in}
    \caption{Top - Original and warped signal waveform and spectrogram on a log time axis. Bottom - Estimated modes with warping (red circles) and without  warping (blue diamonds). Left - modes in z-plane with direct estimation. Right - modes in z-plane after unwarping. The plots are zoomed in up to a polar angle of $\frac{\pi}{12}$ rad ($1837.5$~Hz) to show the lower frequency modes, and grey lines are at harmonics of the fundamental.}
    \label{fig:ir}
\end{figure*}

As an alternative to the subband processing that is done in FZ-ARMA and FZ-ESPRIT, we propose Frequency-Warped ESPRIT (FW-ESPRIT). FW-ESPRIT warps the second order modal filters such that their poles become more damped and further apart in frequency. Frequency warping \cite{harma2000frequency} warps a uniform frequency axis to a non-uniform one. Warping is done by replacing the unit delay $z^{-1}$ with a first-order allpass filter $\zeta^{-1}$, given by:
\begin{equation}
\label{eq:ap}
    \zeta^{-1} = \frac{z^{-1} -\rho}{1-\rho z^{-1}}
\end{equation}
where $\rho$ is the warping factor. Figure~\ref{fig:map} shows the mapping between natural frequencies, $\omega$, and warped frequencies, $\tilde{\omega}$, determined by the phase of the allpass filter in Eq.~(\ref{eq:ap}). The negative group delay of this allpass filter gives the slope of the mapping as a function of frequency,
\begin{equation}
\begin{split}
\label{eq:grp_del}
    \tilde{\omega} &=  \arctan{\left(\frac{(1 - \rho^2) \sin{\omega}}{(1+\rho^2)\cos{\omega} - 2\rho}\right)} \\
    \frac{d\tilde{\omega}}{d\omega} &= -\frac{1-\rho^2}{1+\rho^2-2\rho\cos{\omega}}.
\end{split}
\end{equation}

By choosing an appropriate warping factor, $\rho^*$, we can map uniformly spaced points on the frequency axis to a non-uniformly spaced warped frequency axis, such as approximating a Bark scale, as shown in Fig.~\ref{fig:map}. For a sampling frequency of $f_s$~Hz, the optimal warping factor for Bark mapping is given by
\begin{equation}
\label{eq:bark}
    \rho^* = 1.0674\left[\frac{2}{\pi} \arctan{(0.6583f_s)} \right]^{0.5} - 0.1916
\end{equation}
which approximately equals $0.75$ for $f_s = 44.1$~kHz, \cite{smith1999bark}. 

For modes near DC, warping increases the frequency by a factor of $(1+\rho)/(1-\rho)$. High-frequency modes near the band-edge get compressed by a factor of $(1-\rho)/(1+\rho)$. Similarly, the decay times of the low-frequency modes are decreased by a factor of $(1+\rho)/(1-\rho)$, and that of the high-frequency modes are increased by a factor of $(1-\rho)/(1+\rho)$. The choice of the warping factor should depend on the application, and how much frequency zooming is desired. For a zooming factor of $K_z$ in the lower frequencies, the warping factor can be selected as $\rho = \frac{K_z-1}{K_z+1}$. 

To warp a signal or an impulse response, $h(t)$, the filter structure in Fig.~\ref{fig:tikz_warp} needs to be implemented. As a result of the warping, the signal frequencies have non-uniform group delay. The original and warped piano signals with $\rho = \rho^*$ are shown in Fig.~\ref{subfig:ir}, \ref{subfig:irwarp}. As visible in the frequency domain, the modes in the warped response have been shifted to higher and further spaced frequencies, and their dampings have been increased (approximately by a factor of $7$). 

To estimate the modes, we first warp the response, estimate the warped modal parameters, and unwarp the frequencies and dampings before estimating the amplitudes using the unwarped impulse response. The measured signal is warped with a warping factor of $\rho^*$ (Eq.~\ref{eq:bark}) and its modes are estimated using the method described in Section~\ref{sec:modes}. 

Modal estimation with frequency warping overlooks some high frequency modes because of compression of the frequency axis near the band-edge. These modes are required for accurate reconstruction of the transient. As a workaround, two sets of modes are calculated from the signal, one with and one without frequency warping. A combined set of modes is used, with lower frequency modes from the warped set and higher frequency modes from the non-warped set. A frequency cutoff threshold, $\omega_c$, is selected for combining the modes by setting the slope of the mapping to $1$, where the warped frequency is equal to the non-warped frequency. This gives $\omega_c = \cos^{-1}{(\rho)}$. To unwarp the mode frequencies and dampings, the poles are unwarped, and Eq.~(\ref{eq:log_eig}) is used to get the mode frequencies and dampings from the unwarped poles.
\begin{equation}
    \begin{split}
        \tilde{\psi}_m &= e^{\left(j\tilde{\omega}_m - \tilde{\alpha}_m\right)} \\
        \psi_m &= \frac{-\rho + \tilde{\psi}_m^*}{1 - \rho \tilde{\psi}_m^*}. \\
    \end{split}
\end{equation}

The modes estimated from the three coupled strings of a measured piano note at $220$~Hz, without and with frequency warping are plotted as poles on the z-plane in Figs.~\ref{subfig:no_warp_poles},~\ref{subfig:warp_poles} respectively. Beating modes can be clearly seen in Fig.~\ref{subfig:warp_poles}, where overlapping circles in the zoomed plot indicate triplets caused by the three coupled strings vibrating together. Low-frequency beating modes are captured by frequency warping, which are otherwise missed in the non-warped case. 
% However, due to reasons explained, frequency warping misses some of the high frequency modes. 

%%%%%%%%%%%%%%%%%%%%%%%%%%%%%%%%%%%%%%%%%%%%%%%%%%%%%%%%%%%%%%
\subsection{Mode optimization}
\label{sec:opt}

To fine-tune the estimated modes, we propose a new time-domain iterative mode optimization method. 
The modeled signal, $\hat{h}(t)$, Eq.~(\ref{eq:modes}) can also be written as
\begin{equation}
    \label{eq:diff_modes}
    \hat{h}(t) = \sum_{m=1}^M e^{-\alpha_m t} \left[ \gamma_{s_m} \sin{(\omega_m t)} + \gamma_{c_m} \cos{(\omega_m t)} \right] 
\end{equation}
\noindent where $\gamma_{s_m}, \gamma_{c_m}$ are the real mode amplitudes associated with the sine and cosine components of the $m$th mode. Using Eq.~(\ref{eq:diff_modes}) has the advantage of making the cost function real. In vector form, Eq.~(\ref{eq:diff_modes}) can be written as
\begin{equation}
    \begin{split}
    \hat{\bm{h}} &= 
    \begin{bmatrix} \Im(\bm{V}) & \Re(\bm{V}) \end{bmatrix} 
    \begin{bmatrix} \bm{\gamma_s} \\
    \bm{\gamma_c}
    \end{bmatrix}
    \\
    &= 
    \begin{bmatrix}
    e^{-\alpha_1 0}\sin{(\omega_1 0)} & \cdots   & e^{-\alpha_M 0}\cos{(\omega_M 0)} \\ \vdots & \cdots & \vdots \\
    \vdots & \cdots & \vdots \\
    \vdots & \cdots & \vdots \\
    e^{-\alpha_1 T}\sin{(\omega_1 T)} & \cdots  &  e^{-\alpha_M T}\cos{(\omega_M T)}
    \end{bmatrix}
    % \begin{bmatrix}
    % e^{-\alpha_1 0}Section~in{(\omega_1 0)} & \cdots   & e^{-\alpha_M 0}Section~in{(\omega_M 0)} & e^{-\alpha_1 0}\cos{(\omega_1 0)} & \cdots &  e^{-\alpha_M 0}\cos{(\omega_M 0)} \\
    % \vdots & \vdots & \ddots &\ddots & \vdots & \vdots \\
    % e^{-\alpha_1 T}Section~in{(\omega_1 T)} & \cdots  & e^{-\alpha_M T}Section~in{(\omega_M T)} & e^{-\alpha_1 T}\cos{(\omega_1 T)} & \cdots &  e^{-\alpha_M T}\cos{(\omega_M T)}
    % \end{bmatrix}
    \begin{bmatrix}
    \gamma_{s_1} \\
    \vdots \\
    \gamma_{s_M} \\
    \gamma_{c_1} \\
    \vdots \\
    \gamma_{c_M}
    \end{bmatrix} 
    \end{split}
\end{equation}
% \normalsize
We can now form a vector of mode dampings and frequencies, $\theta = [\bm{\alpha}^T \ \bm{\omega}^T]$ to optimize. The time-domain non-linear cost function and its gradient are 
\begin{equation}
    \begin{split}
        J(\theta) &= \frac{1}{2}||\bm{h} - \hat{\bm{h}}(\theta)||^2_2 \\
        &= \frac{1}{2} \left(\bm{h} - \hat{\bm{h}}(\theta)\right)^\top \left(\bm{h} - \hat{\bm{h}}(\theta)\right) \\
        \nabla_\theta J(\theta) &= D_\theta (\hat{\bm{h}}(\theta))^\top \left[\hat{\bm{h}}(\theta) - \bm{h}\right]
    \end{split}
\end{equation}
where $\bm{h}$ is the measured signal vector, $\hat{\bm{h}}(\theta)$ is the modeled signal vector and $D_\theta(\hat{\bm{h}}(\theta))  \in \mathbb{R}^{T \times 2M}$ is the Jacobian matrix of $\hat{\bm{h}}(\theta)$ with respect to $\theta$. Each entry of the Jacobian matrix is given by
\small{
\begin{equation}
    \begin{split}
        D_\theta(t,m) &= \frac{\partial \hat{h}(t)}{\partial \alpha_m} \\ &= 
        -te^{-\alpha_m t} \left[ \gamma_{s_m} \sin{(\omega_m t)} +  \gamma_{c_m} \cos{(\omega_m t)} \right]\\
         D_\theta(t,2m) &= \frac{\partial \hat{h}(t)}{\partial \omega_m} \\ &= 
         te^{-\alpha_m t} \left[ \gamma_{s_m} \cos{(\omega_m t)} -  \gamma_{c_m} \sin{(\omega_m t)} \right]
    \end{split}
\end{equation}
}
\normalsize
For faster computation, we bandpass filter the signal into several frequency bands and optimize the modes in each band. We use MATLAB's \texttt{fmincon} optimizer with initial mode parameter estimates calculated from Section~\ref{sec:modes}. The amplitudes $\gamma_{s_m}, \gamma_{c_m}$ are re-calculated in each iteration of the optimization using least squares. 

Providing \texttt{fmincon} with the gradient of the cost function speeds up computation significantly. We also set constraints such that $\omega_{0_m} - \delta_\omega < \omega_m < \omega_{0_m} + \delta_\omega$ and $\alpha_{0_m} - \delta_\alpha < \alpha_m < \alpha_{0_m} + \delta_\alpha$, where $\omega_0$ and $\alpha_0$ are the initial estimates and $\delta_\omega, \delta_\alpha$ are the maximum deviations allowed from the initial point.  We ensure the mode frequencies are ordered in ascending order, i.e., $\omega_{m-1} \leq \omega_m \leq \omega_{m+1}$.

The cost function is non-convex, so converging to a local minimum is probable. To avoid this, the initial estimate has to be close to the global optimum. We run the optimizer for a maximum of $500$ function counts and convergence is achieved if the error is small enough ($ \leq 1e-4$), or the step size calculated by \texttt{fmincon} is small enough ($ \leq 1e-9$).

\begin{algorithm}
\caption{Mode optimization pseudocode}
\begin{algorithmic}
\REQUIRE $\alpha_{b,0} - \delta_\alpha \leq \theta_{\text{1:M}} \leq \alpha_{b,0} + \delta_\alpha$,
    $\omega_{0,b} - \delta_\omega \leq
    \theta_{\text{M+1:2M}} \leq \omega_{b,0} + \delta_\omega \ \forall \ b$ 
    
\FOR {$b = 1, \cdots, N_b$}

    \STATE $\theta_{0} = [\alpha_{b,0} \ \ \omega_{b,0}]^\top$
    
    \REPEAT 
    \STATE $\hat{\bm{h}}(\theta_i) \leftarrow   \begin{bmatrix}
    \Im(\bm{V}(\theta_i) & \Re(\bm{V}(\theta_i)  \end{bmatrix} 
    \begin{bmatrix} \bm{\gamma_s}\\
    \bm{\gamma_c} \end{bmatrix}$

    \STATE $J(\theta_i) \leftarrow 0.5 \ ||\bm{h} - \hat{\bm{h}}(\theta_i)||^2 $

    \STATE  $\theta_{i+1} \leftarrow \arg \min_\theta J(\theta_i)$

    \STATE 
        $ [\bm{\gamma_s} \ \
        \bm{\gamma_c}
        ]^\top \leftarrow  \begin{bmatrix} \Im(\bm{V}(\theta_{i+1}) & \Re(\bm{V}(\theta_{i+1}) \end{bmatrix} ^\dagger \bm{h}$
        
    \UNTIL convergence
    
    \STATE $\alpha^*_b, \omega^*_b = \theta^*_i$
    
\ENDFOR
\end{algorithmic}
\end{algorithm}

%%%%%%%%%%%%%%%%%%%%%%%%%%%%%%%%%%%%%%%%%%%%%%%%%

\section{Applications}
\label{sec:appli}

%%%%%%%%%%%%%%%%%%%%%%%%%%%%%%%%%%%%%%%%%%%%%%%%
\subsection{Coupled Piano Strings}
\label{ssec:coupled_piano}

\begin{figure*}[t]
\centering
\subfloat[]{\includegraphics[height=0.2\textheight]{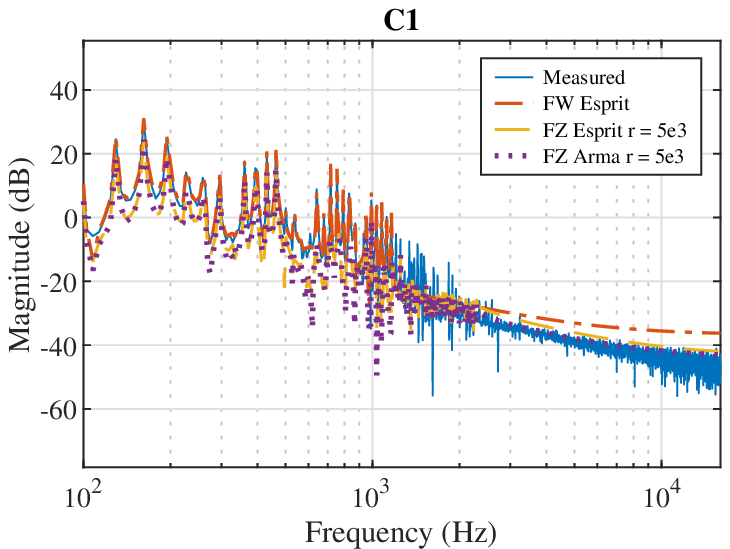} \label{subfig:C1}}
\qquad \qquad \qquad
\subfloat[]{\includegraphics[height=0.2\textheight]{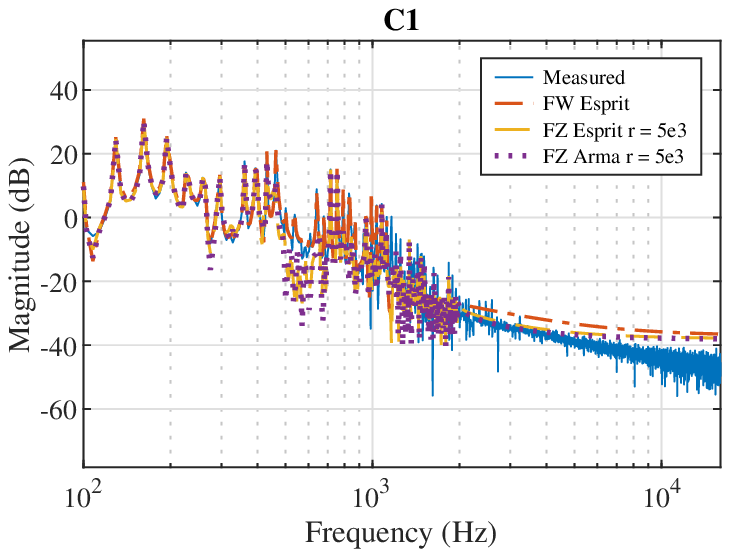} \label{subfig:C1_opt}} 
\\ 
\subfloat[]{\includegraphics[height=0.2\textheight]{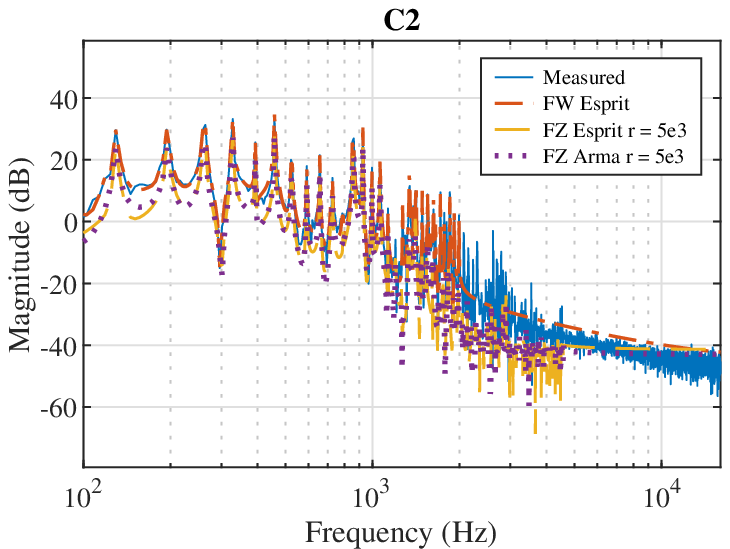} \label{subfig:C2}}
\qquad \qquad \qquad
\subfloat[]{\includegraphics[height=0.2\textheight]{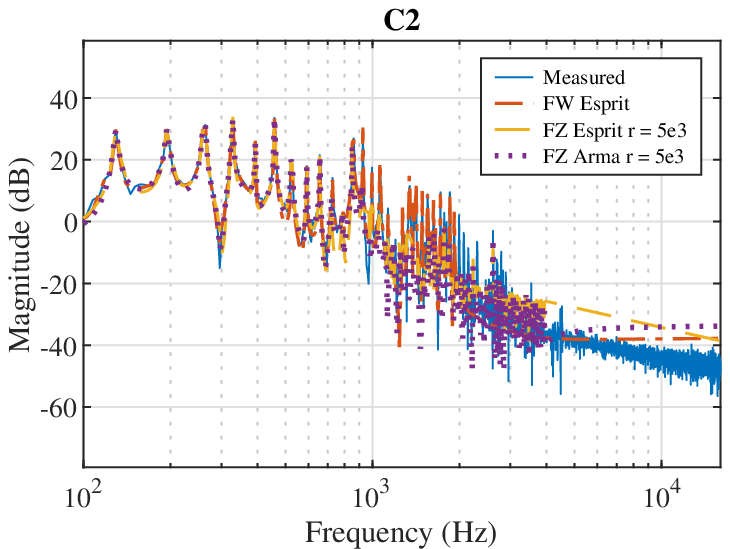} \label{subfig:C2_opt}}
\\ 
\subfloat[]{\includegraphics[height=0.2\textheight]{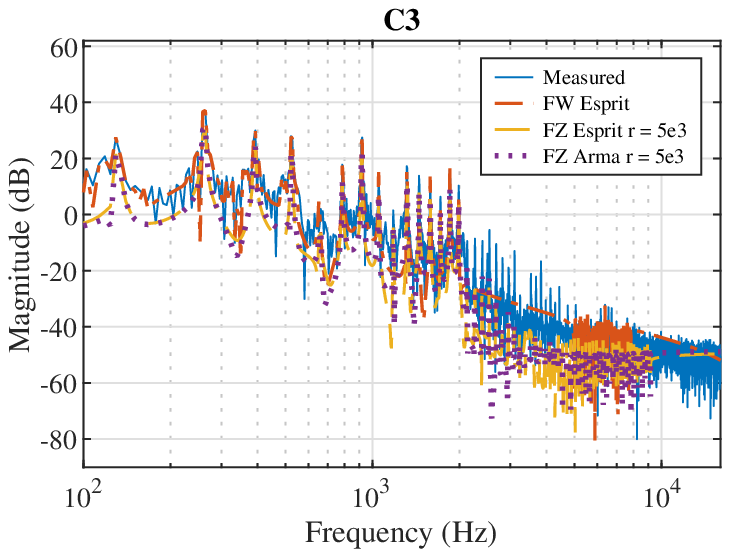} \label{subfig:C3}}
\qquad \qquad \qquad
\subfloat[]{\includegraphics[height=0.2\textheight]{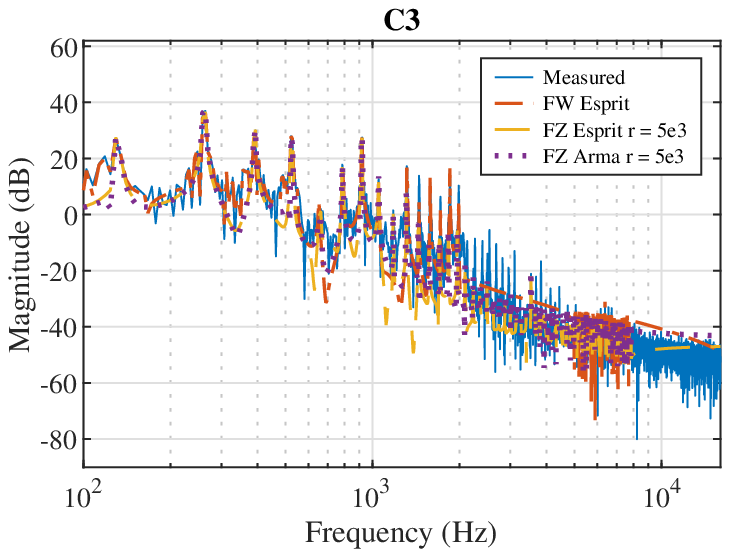} \label{subfig:C3_opt}}
\\ 
\subfloat[]{\includegraphics[height=0.2\textheight]{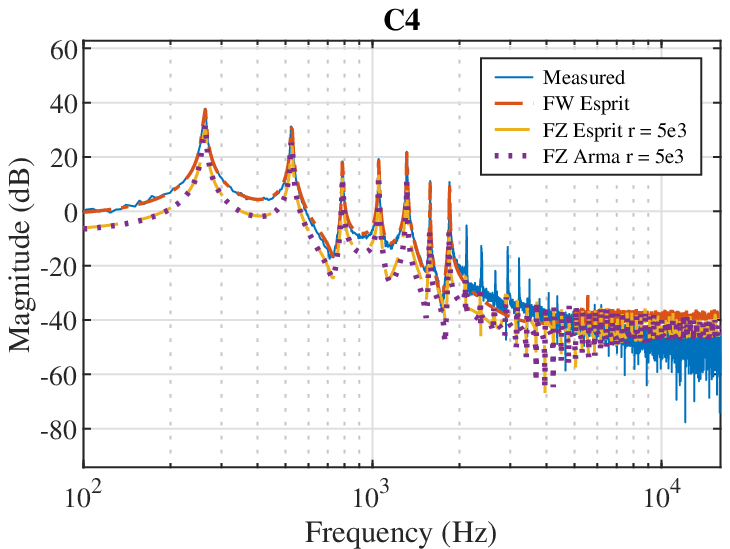} \label{subfig:C4}}
\qquad \qquad \qquad
\subfloat[]{\includegraphics[height=0.2\textheight]{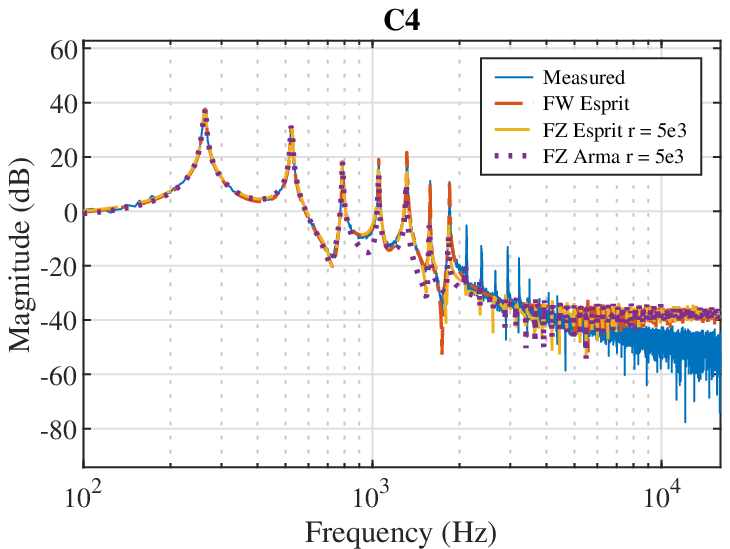} \label{subfig:C4_opt}}
\caption{Frequency response of measured and synthesized notes. Left - without mode optimization, Right - with mode optimization.}
\label{fig:freq_response}
\end{figure*}

Coupled string systems typically have two or more strings coupled at a non-rigid bridge.  The eigenanalysis of two strings coupled at a bridge is explained in \cite{smith2010physical}.  The piano hammer striking the strings produces motion primarily in the transverse direction, but also in the longitudinal direction. The transverse vibrations are much stronger initially, but attenuate quickly whereas the weaker longitudinal vibrations persist. There is significant coupling between transverse and longitudinal vibrations \cite{bank2005generation}. Transverse vibrations produce modes at harmonic frequencies. The longitudinal motion is continuously excited by the transverse vibration along the string. The forced response to this excitation gives rise to phantom partials, while the free response produces the components corresponding to the longitudinal modal frequencies.  The \textit{aftersound}, or the second stage of a two-stage decay is caused by a combined effect of string coupling and shift from transverse to longitudinal polarization. 
% There are two normal modes of vibration - the ``in-phase'' vibrations see a longer effective string length, and move the bridge vertically, causing a more rapid decay, whereas ``anti-phase'' vibrations see no length correction because the bridge is rigid in this case. The lengthening of the string due to ``in-phase'' vibrations makes the piano attack decay faster and flattens it, whereas the \textit{aftersound}, composed of ``anti-phase'' vibrations, decays slowly and remains in tune. 

Coupled piano strings are slightly mistuned, and the amount of mistuning either causes audible beating (amplitude modulation), or a beatless \textit{aftersound}. Coupled strings vibrate in phase immediately after impact, producing the \textit{prompt} sound. Because their frequencies are slightly different, they become out of phase eventually. Once this phase offset becomes approximately a half period of one of the frequencies, the movement at the bridge almost completely cancels, and sound is sustained. Modal parameters of coupled strings can be used in conjunction with other physical models, such as digital waveguides, for more realistic sound synthesis. A hybrid modal-waveguide model has been used to synthesize coupled strings in \cite{lee2009analysis}. Coupled digital waveguides have been used in \cite{aramaki2001resynthesis}.

\begin{table*}
\caption{Mode estimation results without optimization}
\centering
\begin{tabular}{l|lll|lll|lll}
\hline
 Note & \multicolumn{3}{l|}{FW-ESPRIT} & \multicolumn{3}{l|}{FZ-ESPRIT r = 5000} &  \multicolumn{3}{l}{FZ-ARMA r = 5000} \\[5pt] 
\hline
 & MSE (dB) & Time (s) & M & MSE (dB) & Time (s) & M & MSE (dB) & Time (s) & M  \\ [2pt]
 \hline
C1		& -118.25 &	13.43   & 61  &-101.07 & 21.16 & 290     & -101.11	&35.87	& 666\\
C2		& -94.97  &	27.52	& 119   &-94.16 & 28.13	& 528     & -94.35  &	37.51	& 720\\
C3      & -106.95 & 25.61	& 349   &-91.49 & 42.56 &	628     & -91.63   & 38.17	& 720\\
C4      & -113.47 &	83.24	& 1005  &-102.98 &	49.65 &	769     & -102.91 &	32.33	& 720
\\
\hline
\end{tabular}
\label{table:res_no_opt}
\end{table*}

\begin{table*}
\caption{Mode estimation results with optimization}
\centering
\begin{tabular}{l|lll|lll|lll}
\hline
 Note & \multicolumn{3}{l|}{FW-ESPRIT opt} & \multicolumn{3}{l|}{FZ-ESPRIT opt r = 5000} &  \multicolumn{3}{l}{FZ-ARMA opt r = 5000} \\[5pt] 
\hline
 & MSE (dB) & Time (s) & Iter & MSE (dB) & Time (s) & Iter & MSE (dB) & Time (s) & Iter  \\ [2pt]
 \hline
C1		&-119.52   &20.34	& 93    &-110.52	&13.85 & 1 	&-110.71   &14.63   &1
\\
C2		&-95.04    &69.14   & 424	&-93.59	    & 30.83 & 154   &-95.51	   &57.38   & 214\\
C3      &-108.66   &96.13	& 305	&-95.79  	& 26.81 & 142   &-99.91	   &48.57   & 125\\
C4      &-111.44   &48.22	& 118   &-116.77	& 23.99 & 74    &-112.04   &41.72  & 136	
\\
\hline
\end{tabular}
\label{table:res_opt}
\end{table*}

We evaluated FW-ESPRIT against FZ-ESPRIT and FZ-ARMA on the MIS\footnote{http://theremin.music.uiowa.edu/MISpiano.html} dataset of piano recordings for the C note spanning four octaves, from $32.7$~Hz (C$1$) to $261.6$~Hz (C$4$)\footnote{Sound examples are available at \url{https://ccrma.stanford.edu/~orchi/Modal/warped_modal.html}.}. For FZ-ESPRIT, we warp the measured signal with a warping coefficient $\rho^*=-0.77$ (Eq.~\ref{eq:bark}) at a sampling rate of $f_s = 44.1$~kHz. The order of the Hankel matrix used is $\mathbb{R}^{T\times T}, T = 2048$. A large order gives more accurate mode estimates at the cost of computational complexity. To determine the number of modes, $M$, we order the singular values by magnitude and find its \textit{knee-point} \cite{kneepoint}. 

For FZ-ESPRIT and FZ-ARMA, we center the bandpass filters around frequency $f_n$~Hz after taking into account string stiffness \cite{young1952inharmonicity} with an inharmonicity coefficient of $B = 10^{-4}$ and a fundamental frequency of $f_0$~Hz,
\begin{equation}
    f_n = n f_0  \sqrt{1 + Bn^2}
\end{equation}
 A fourth order Butterworth filter with a bandwidth of $1/10$th the fundamental frequency is used. The downsampling factor is selected to be $K_z = 5000$. The ARMA model order is fixed to be $12$, as suggested in \cite{karjalainen2002frequency}. For all methods, modes around the first $60$ partials are estimated. 

The time domain mean-squared errors without optimization are shown in Table~\ref{table:res_no_opt}. FW-ESPRIT outperforms the other two methods at a comparable run-time. For the ESPRIT based methods, computation time is directly proportional to the number of estimated modes. As the fundamental frequency increases, so does the number of modes. FZ-ARMA has mean-squared errors  comparable to those of FZ-ESPRIT, since they essentially perform the same operations of downsampling, filtering and estimating. The ESPRIT based algorithms have an advantage over FZ-ARMA, since the detection of the number of modes is automatic. 
% The frequency spectrum of the measured and modeled signals are shown in Figs.~\ref{subfig:C1}, \ref{subfig:C2}, \ref{subfig:C3}, \ref{subfig:C4}.

The results of the proposed post-processing optimization scheme are shown in Table~\ref{table:res_opt}. Optimization parameters were selected to be $\delta_\alpha = 0.1\alpha_0$ and $\delta_\omega = 0.5$~Hz (window of $1$~Hz around estimated mode frequencies). The third column of the table denotes the number of iterations to convergence. The total run-time of the optimization depends on this, as well as the number of modes being optimized. In general, optimization improves the mean-squared error, except in the case of C$4$ estimation with FW-ESPRIT. The computation time is reasonable for offline applications. The frequency spectrum of the measured and modeled signals with and without optimization are shown in Fig~\ref{fig:freq_response}. Fits to the measured data is improved with optimization.

%%%%%%%%%%%%%%%%%%%%%%%%%%%%%%%%%%%%%%%%%%%
\subsection{Room Impulse Response}
\label{ssec:rir}

\begin{figure*}[!ht]
\centering
    \subfloat[FW-ESPRIT, $M = 2994$]{
    \includegraphics[height=0.35\textheight]{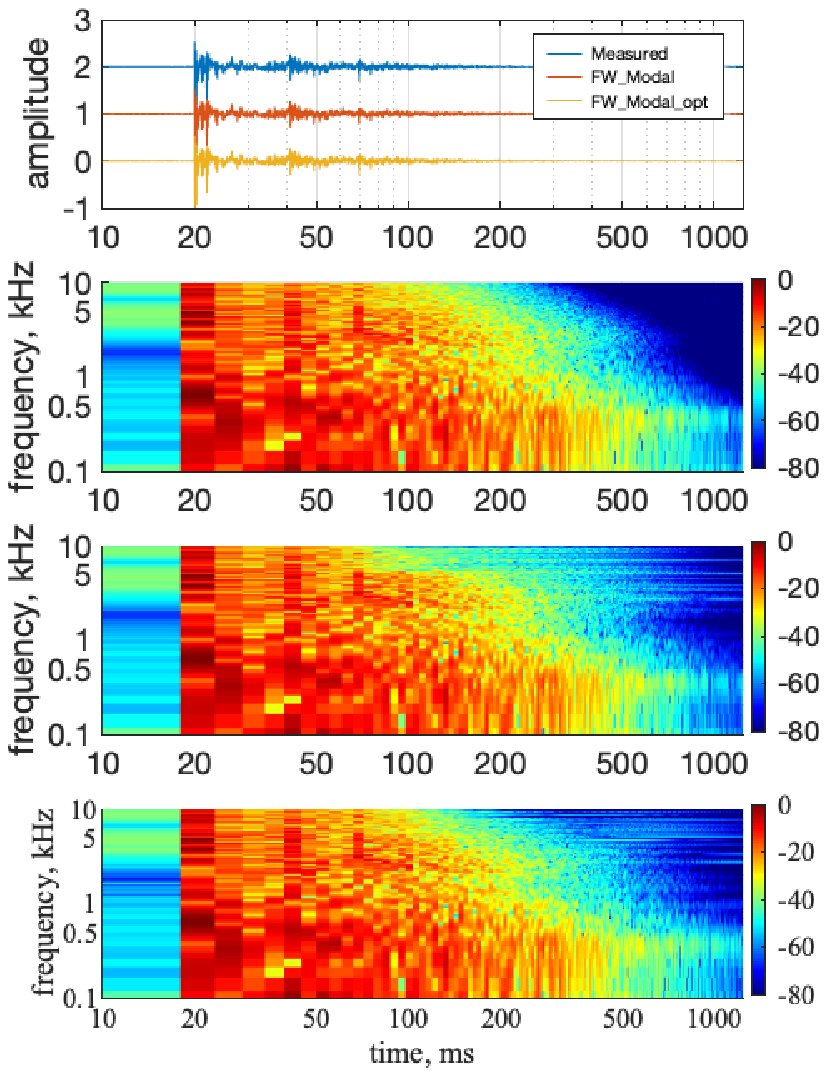} 
    \label{subfig:rir_fw_esprit}}\qquad
    \subfloat[FW-ESPRIT, $M=1402$]{
    \includegraphics[height=0.35\textheight]{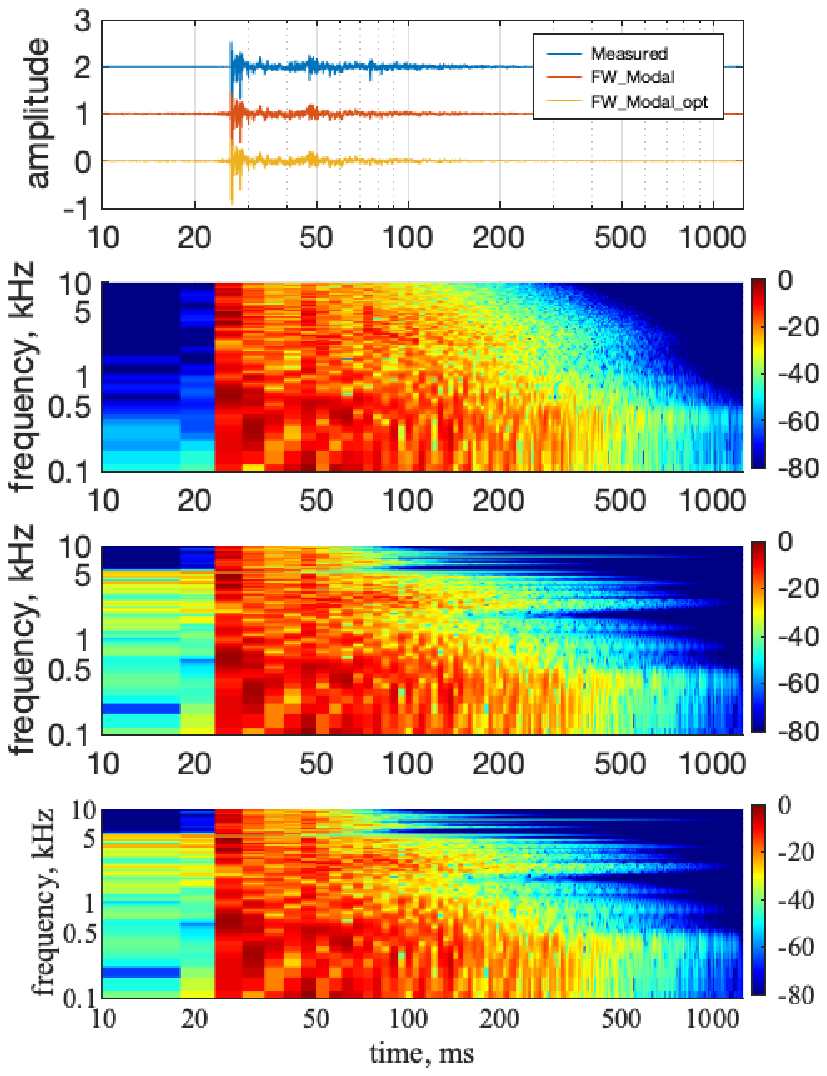} 
    \label{subfig:rir_fw_esprit_alt}} \\
    \subfloat[FZ-ESPRIT, $M=1421$]{
    \includegraphics[height=0.35\textheight]{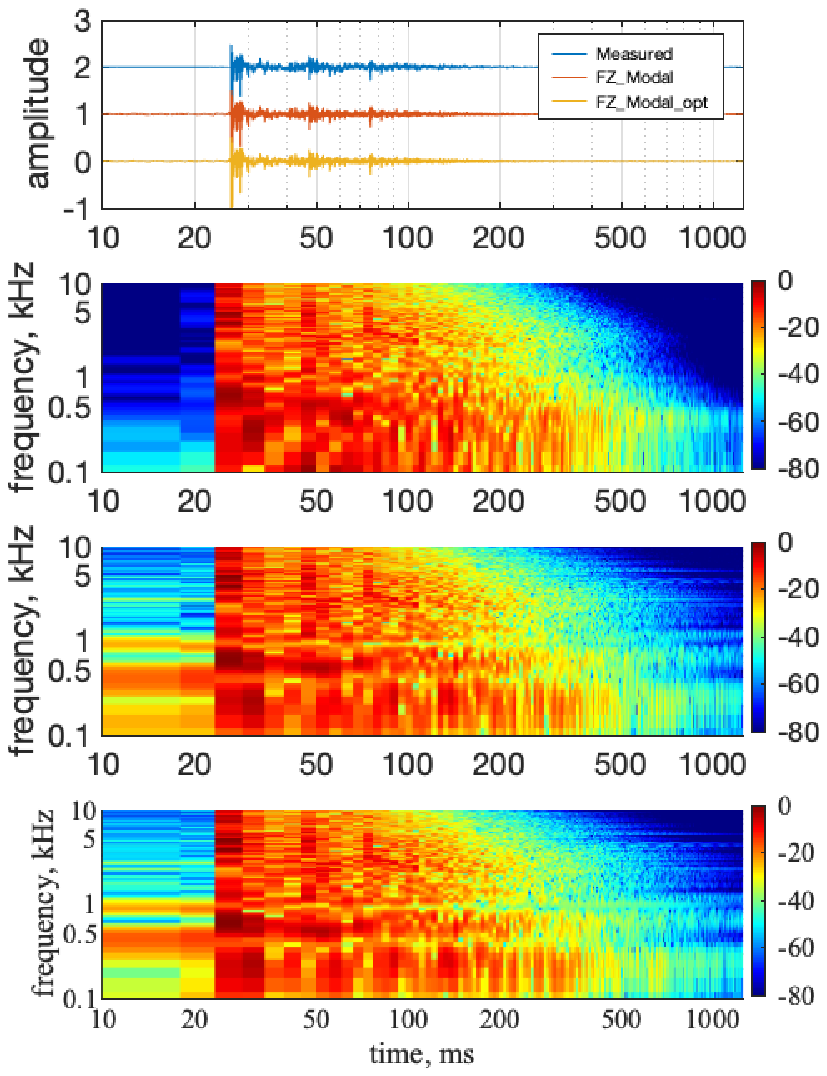} 
    \label{subfig:rir_fz_esprit}} \qquad
    \subfloat[FZ-ARMA, $M=1368$]{
    \includegraphics[height=0.35\textheight]{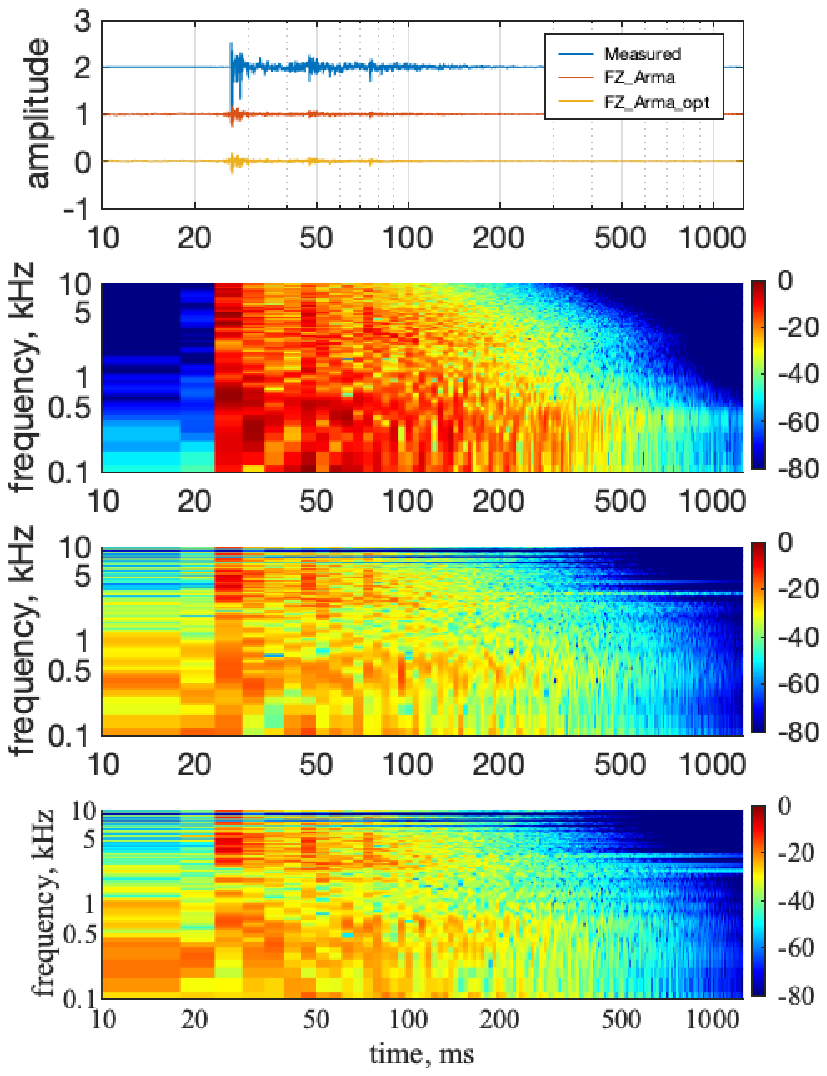} 
    \label{subfig:rir_fz_arma}}
    \caption{RIR modeling results. Spectrograms from top - Measured, Modeled without optimization, Modeled with optimization.}
\end{figure*}

\noindent The sound field in a room is given by the 3D wave equation \cite{kuttruff2016room},
\begin{equation}
\frac{\partial^2p}{\partial t^2} = c^2 \left( \frac{\partial^2p}{\partial x^2} + \frac{\partial^2p}{\partial y^2} + \frac{\partial^2p}{\partial z^2} \right)
\end{equation}
where $p$ is the acoustic pressure, $c$ is the speed of sound in the medium, and $x,y,z$ are Cartesian coordinates. The time-independent form of the wave equation in the Fourier domain is known as the Helmholtz equation, characterized by the wave-number $k$,
\begin{equation}
    \label{eq:helmholtz}
     \frac{\partial^2p}{\partial x^2} + \frac{\partial^2p}{\partial y^2} + \frac{\partial^2p}{\partial z^2} + k^2 p= 0
\end{equation}

The solutions to this equation are standing waves, or room modes. For a rectangular room with rigid walls and dimensions $L_x, L_y$ and $L_z$ in directions $x,y,z$ respectively, the wave numbers, $k$, and mode frequencies, $f$, are given by
\begin{equation}
    \begin{split}
        k^2 &= k_x^2 + k_y^2 + k_z^2; \;  f = \frac{c}{2 \pi} k \\
        k_\mu &= \frac{n_\mu \pi}{L_\mu}; \ \mu \in (x,y,z), \ n_\mu \in \mathbb{Z}^+ 
    \end{split}
\end{equation}
For varying values of $n_x,n_y$ and $n_z$, a large number of axial, tangential and oblique modes are generated. 

A room impulse response can be characterized by a sum over $M$ such modes (\ref{eq:modes}). The complex amplitudes, $\gamma_m$, are functions of space, determined by the source and receiver positions, whereas frequencies and dampings, $\omega_m$ and $\alpha_m$, are properties of the room and its material. The idea of a modal reverberator is to implement the room modes using separate resonant filters, each driven by the source signal and summed in a parallel structure to form the output \cite{abel2014modal}. 

A medium sized classroom was measured and modeled with FW-ESPRIT, FZ-ESPRIT and FZ-ARMA. The Hankel matrix size for FW-ESPRIT was doubled to $4096$~samples. The non-warped impulse response was multiplied with a damped exponential, $r^n = \exp{(-1.5 n/f_s)}$, before applying ESPRIT to find its modes. This introduced some extra damping to the high frequency modes, i.e., it shrunk the pole radius. Once the modes were estimated, they were undamped by multiplying with $\frac{1}{r}$. After this, the two sets of modes from the warped and non-warped RIR were combined. 

For one experiment, we selected the number of modes by finding the \textit{knee-point} of the singular value curve of the Hankel matrix. This resulted in a good fit with $M = 2994$ modes, as shown in Fig.~\ref{subfig:rir_fw_esprit}. The decay rates of some high frequency modes are overestimated, but the optimization helps take care of that. For another experiment, the number of modes was chosen by finding the singular values whose magnitude was greater than $-18$~dB. This resulted in an SNR of $-18$~dB and $M = 1402$ modes. The results, shown in Fig.~\ref{subfig:rir_fw_esprit_alt}, indicate a poorer fit with colored reverberation.

For FZ-ARMA and FZ-ESPRIT, the processing was done on a non-uniform filterbank approximating the Bark scale, with smaller bandwidth low frequency bands and longer bandwidth high frequency bands. There were $N_b = 20$ frequency bands for FZ-ESPRIT, and $N_b = 40$ bands for FZ-ARMA whose centers were uniformly distributed from $[0, \frac{f_s}{2}]$~Hz, with $f_s  = 48$~kHz, and then warped according to Eq.~(\ref{eq:grp_del}). The filter bandwidth was selected as $\beta_b = 1.2*((f_{b+1} - f_b)/2)$~Hz. A $7$th order Elliptic filter with a stop-band ripple of $-80$~dB, pass-band ripple of $1.5$~dB and cutoff of $\beta_b$~Hz was used. The results of FZ-ESPRIT are shown in Fig.~\ref{subfig:rir_fz_esprit}.

For FZ-ESPRIT, a maximum number of modes, $M_{b_{\text{max}}} = 100$, was allotted per frequency band. If the mode budget was not exhausted in any band, the remaining mode budget was distributed equally among the other bands. This ensured that higher frequency bands with a greater bandwidth (which contain many rapidly decaying modes) had an adequate mode budget. The downsampling factor was $K_z=8$. 

For FZ-ARMA, the maximum allowed zero and pole order was set to $N_{\text{max}} = 30$, $P_{\text{max}} = 150$ respectively. The pole and zero order of each band was selected by a non-linear mapping of the bandwidth, 
\begin{equation}
\begin{split}
  \beta_{bn} &= \frac{\beta_b}{\max_{b = 1,\cdots,N_b} \beta_b}, \ \ \beta_{bn} \in (0,1]\\
    \{N,P\}_b &= \nint[\Big]{\frac{\tanh{(3\beta_{bn})}}{\tanh{(3)}} \times \{N,P\}_{\text{max}}}
    \end{split}
\end{equation}
where $\nint{}$ denotes rounding operation to the nearest integer. This also ensures that the modes are distributed according to the filter bandwidths. The downsampling factor was selected to be $K_z=100$. For all methods, the optimization window for the frequency estimates was set to $\delta_\omega = 2$~Hz, and a maximum of $100$ iterations were run for each frequency band. The results are shown in Fig.~\ref{subfig:rir_fz_arma}.

It is evident from the plots and the sound examples that FZ-ARMA fails at modeling the RIR. With the right selection of parameters, it can potentially model a certain narrow frequency range accurately, as discussed in \cite{karjalainen2002frequency}. Between FZ-ESPRIT and FW-ESPRIT, the former gives perceptually impressive results with almost half the number of modes.

%%%%%%%%%%%%%%%%%%%%%%%%%%%%%%%%%%%%%%%%%%
\section{Discussion}
\label{sec:disc}
\begin{itemize}
    \item Compared to FZ-ARMA, the ESPRIT based methods have the advantage of automatic order selection. The performance of ARMA modeling depends on the model orders selected, which needs to be done manually on a case-by-case basis. In the ESPRIT based methods, we algorithmically look at the magnitudes of the singular values of the Hankel matrix to make this decision.
    
    \item There are a few options for selecting the number of modes. One option is to look for the \textit{knee-point} in the singular value curve, which indicates the point after which the curve drops off. This produces the best fit at the cost of picking a large number of modes. The other option is to choose an SNR for each frequency band, and pick the singular values whose magnitudes lie above the SNR. This method gives a constant SNR in each frequency band.
    
    \item FZ-ARMA needs a large downsampling factor to give reasonable results. A large downsampling factor can easily miss the fast decaying modes. 
    % For a very large downsampling factor, one might as well detect peaks in the FFT to find the modes, instead of doing ARMA modeling.
    
    \item In FZ-ESPRIT, one has to manually select the downsampling factor, number of frequency bands, transition frequencies and bandwidths. The only parameter that needs tuning in FW-ESPRIT is the warping coefficient. 
    \item FZ-ARMA is inept at modeling RIR modes. FZ-ESPRIT produces the best perceptual results with fewer modes than FW-ESPRIT. FZ-ESPRIT is preferred for modeling RIRs, whereas FW-ESPRIT is ideal for modeling harmonic signals, such as those of musical instruments. 
    All methods have been implemented in a modal estimation toolbox \cite{modal-estimation-toolbox}.
    
\end{itemize}

%%%%%%%%%%%%%%%%%%%%%%%%%%%%%%%%%%%%%%%%%%
\section{Conclusion}
\label{sec:conclusion}

A modal estimation technique has been proposed, which operates on a warped frequency axis, and is well-suited to modeling low-frequency beating modes. Frequency warping gives higher resolution at low frequencies and increases damping of the modes. This method achieves results comparable to subband modal estimation (FZ-ESPRIT) with a  computational advantage, and gets better fits than FZ-ARMA. To fine-tune the estimated mode parameters, the error between our model and the measured response has been optimized. 

Two applications of the proposed method were demonstrated --- modeling of coupled piano strings and room impulse responses. Subband filter design and model order selection was discussed in detail for all the methods. The results of FW-ESPRIT, compared with FZ-ESPRIT and FZ-ARMA, show the efficacy of the proposed method in modeling linear systems with densely distributed modes with minimal parameter tuning.

%%%%%%%%%%%%%%%%%%%%%%%%%%%%%%%%%%%%%%%%%%%%%%%%%%%%%%%%%%%%%%%%%%
\bibliographystyle{IEEEtran}
\bibliography{template}  %%% Uncomment this line and comment out the ``thebibliography'' section below to use the external .bib file (using bibtex) .

\end{document}